\documentclass[aps, pra, preprint, groupedaddress, amsfonts,
               amsmath, amssymb, showpacs, nofootinbib]{revtex4-1}
\usepackage{microtype}
\usepackage{graphicx}
\usepackage{amssymb}
\usepackage{mathtools}
\usepackage{epstopdf}
\usepackage[utf8]{inputenc}
\usepackage[T1]{fontenc}
\usepackage[usenames,dvipsnames]{xcolor}
\usepackage{hyperref}

\newcommand{\bra}[1]{\langle #1|}
\newcommand{\ket}[1]{|#1\rangle}
\newcommand{\braket}[2]{\langle #1|#2\rangle}

\usepackage{color}

\begin{document}
\title{A criterion for an effective discretization of a continuous Schr\"odinger
spectrum using a pseudostate basis}

\author{Tom Kirchner}  
\email[]{tomk@yorku.ca}
\affiliation{Department of Physics and Astronomy, York University, Toronto, Ontario, Canada M3J 1P3}

\author{Marko Horbatsch}  
\email[]{marko@yorku.ca}
\affiliation{Department of Physics and Astronomy, York University, Toronto, Ontario, Canada M3J 1P3}

\date{\today}
\begin{abstract}
We consider a Hamiltonian $\hat H$ with a (partially) continuous spectrum and examine the zero-overlap condition 
which involves the projection onto exact continuum eigenstates of a set of pseudostates obtained from the
diagonalization of $\hat H$ in a finite basis of square-integrable functions. 
For each projected pseudostate the condition implies the occurrence 
of zeros at all energies that correspond to the pseudo-continuum matrix eigenvalues, 
except for the eigenenergy associated with that pseudostate.
This feature was observed for the Coulomb continuum represented in a Laguerre basis 
[M.~McGovern \textit{et al.}, Phys. Rev. A \textbf{79}, 042707 (2009)] and later explained
using special properties of the Laguerre functions
[I.~B.~Abdurakhmanov \textit{et al.}, J. Phys. B \textbf{44}, 075204 (2011)].
We establish that a sufficient condition for the zero-overlap condition to occur is that the 
image space of the operator
$\hat Q \hat H \hat P$, where $\hat P$
is the projection operator onto the subspace spanned by the basis and $\hat Q = \hat 1 - \hat P$ its
complement, has dimension one.
We show that the condition is met 
for the one-dimensional free-particle problem by a basis of harmonic oscillator
eigenstates and by a minimal set of simple momentum-space wave functions, and for the Coulomb problem by a Laguerre basis, thus offering an alternative proof for
the latter case.
The zero-overlap condition ensures that in, e.g., an ionizing collision or laser-atom interaction process, 
transition probabilities obtained from 
the projection of a time-propagated pseudostate-expanded
system wave function onto eigenstates of $ \hat H $ are asymptotically stable.
\end{abstract}

\maketitle
\section{Introduction}
\label{intro}
The discretization of an operator with a (partially) continuous spectrum in terms of a
finite set of square-integrable ($L^2$) basis functions is a standard approach to 
dealing with a variety of physics problems.
The primary motivation for using
such an inherently inexact framework is the computational advantage of working with normalizable $L^2$ functions,
discrete matrix representations and summations, as opposed to continuum functions and 
integral representations. 
Formalisms that are based on this idea 
are manifold, ranging from techniques for the calculation of photoionization cross
sections, autoionization widths and elastic scattering phase shifts to coupled-channel calculations for
inelastic and ionizing light- and heavy-particle collisions.
Some of the relevant original ideas are described in 
Refs.~\cite{heller73, heller74, yamani75, broad78, reading79, bransden84, macias88, stelbovics89}. 
In the context of stationary scattering theory the so-called $J$-matrix method has been reviewed
in Ref.~\cite{jmatrix}.
Relatively recent review-like accounts of stationary and time-dependent methods
are provided in Refs.~\cite{bray17, sca, BTB23, KFR23}.
Laser-matter interactions have been addressed using similar frameworks, see, e.g.,  
Ref.~\cite{uhlmann05} and the recent 
Ref.~\cite{hamer25} in which a hybrid $L^2$ basis and (multi-center) grid method was introduced.

The states obtained from diagonalizing a system Hamiltonian in the space spanned by a set of
$L^2$ basis functions are usually called pseudostates. For an atomic ionization process
the relevant Hamiltonian is usually the undisturbed atomic Hamiltonian, e.g., the Hamiltonian of the
Coulomb problem, if one is interested in the ionization of hydrogen atoms.
An important question arises in this context: How are the pseudostates distributed
over the continuum, i.e., the true eigenstates of that Hamiltonian?
For a pseudostate basis obtained from diagonalizing the Coulomb Hamiltonian in a basis of Laguerre
functions, McGovern~\textit{et al.}~\cite{mcgovern09} found that the pseudostates 
divide the continuum up into ``clumps''.
The clumps overlap, but at each matrix eigenvalue 
the distribution functions for all pseudostates other than the one associated with that eigenvalue
appear to be zero.
McGovern~\textit{et al.} made use of this feature, which we call the zero-overlap
condition, in their time-dependent treatment of differential ionization
of hydrogen (and helium) atoms by antiproton impact.
In a subsequent work by Abdurakhmanov~\textit{et al.}~\cite{ccc1},
a proof for the zero-overlap condition was offered based
on properties of the Laguerre functions previously established in Refs.~\cite{yamani75, stelbovics89}
(see also Ref.~\cite{broad78}).

Recently, it was demonstrated that the zero-overlap condition is a sufficient
condition for the asymptotic stability of ionization probabilities obtained from the projection of a time-propagated
pseudostate-expanded system wave function on the true continuum~\cite{jj26}.
It was noted in an earlier work that this 
obvious requirement for a physical interpretation of the calculated results is in general not
fulfilled if one works with a finite basis representation~\cite{bondarev15}.
It was further shown in Ref.~\cite{jj26} for the Coulomb Hamiltonian that a set of pseudostates obtained from a
basis generator method (BGM) calculation satisfies the zero-overlap condition in good approximation,
indicating that one might be able to understand this feature 
in more general terms.

The purpose of the present paper is to demonstrate that this is indeed the case.
Making use of the Feshbach projector formalism, we establish in Sec.~\ref{sec:theory}
a sufficient criterion for the zero-overlap
condition to hold for any (Hermitian) Hamiltonian with a (partially) continuous spectrum. 
In Secs.~\ref{sec:free} and~\ref{sec:coulomb} we examine the criterion for pseudostate representations
of the one-dimensional free-particle and the three-dimensional Coulomb problem, respectively.
For the latter, we provide an alternative proof for Laguerre basis sets.
A summarizing discussion and conclusions are offered
in Sec.~\ref{sec:conclusions}.
Atomic units, characterized by $\hbar=m_e=e=4\pi\epsilon_0=1$, are used unless otherwise stated.

\section{Zero-overlap criterion}
\label{sec:theory}
Consider the eigenvalue problem of a Hermitian Hamiltonian $\hat H=\hat H^\dagger$ 
\begin{equation}
\hat H \ket{\kappa} = E_{\kappa} \ket{\kappa}  ,
\label{eq:tise}
\end{equation}
whose spectrum is at least in part continuous, in which case we can write $E_{\kappa}=E(\kappa)$.
Note that, depending on the problem investigated, $\kappa$ can be a multiindex. It will be
specified in the applications considered in Secs.~\ref{sec:free} and~\ref{sec:coulomb}.

Following Feshbach~\cite{feshbach62}, we
introduce orthogonal projection operators $\hat P$ and $\hat Q = \hat 1 - \hat P$ 
($\hat Q \hat P = \hat P \hat Q = 0$) and turn Eq.~(\ref{eq:tise}) into a set of
coupled Schr\"odinger equations:
\begin{eqnarray}
	\hat P (\hat H - E_{\kappa} )\hat P \ket{\kappa} &=& -\hat P \hat H \hat Q \ket{\kappa}  ,
	\label{eq:pse} \\
	\hat Q (\hat H - E_{\kappa} )\hat Q \ket{\kappa} &=& -\hat Q \hat H \hat P \ket{\kappa}  .
	\label{eq:qse}
\end{eqnarray}
We further assume that $\cal P$-space is spanned by a set of $N$ orthonormal
Hilbert space vectors:
\begin{equation}
\hat P = \sum_{j=1}^N \ket{\psi_j}\bra{\psi_j} \; , \hskip20pt
\braket{\psi_i}{\psi_j} = \delta_{i,j}. 
\label{eq:pspace1}
\end{equation}
Diagonalizing the Hamiltonian in $\cal P$-space amounts to solving
\begin{equation}
	\hat P (\hat H - \varepsilon_{\ell}) \hat P \ket{\varphi_\ell} =  0 .
	\label{eq:phpse}	
\end{equation}
We can write
\begin{equation}
\hat P \hat H \hat P = \sum_{\ell=1}^N \ket{\varphi_{\ell}} \varepsilon_{\ell} \bra{\varphi_{\ell}}
\label{eq:php}	
\end{equation}
and, as an alternative to Eq.~(\ref{eq:pspace1}), we can use the eigenstates of $\hat P \hat H \hat P$
as a basis for $\cal P$-space:
\begin{equation}
\hat P = \sum_{\ell=1}^N \ket{\varphi_{\ell}}\bra{\varphi_{\ell}} \; , \hskip20pt
\braket{\varphi_{\ell}}{\varphi_{\ell '}} = \delta_{\ell,\ell '}. 
\label{eq:pspace2}
\end{equation}

We are interested in situations in which the projected Schr\"odinger equation~(\ref{eq:pse})
decouples from Eq.~(\ref{eq:qse}). Using~(\ref{eq:php}) and (\ref{eq:pspace2}) one can see that this corresponds to
\begin{equation}
\sum_{\ell=1}^N \ket{\varphi_{\ell}} (\epsilon_{\ell} - E_{\kappa}) \braket{\varphi_{\ell}}{\kappa} = 0 .
\label{eq:zeroo1}
\end{equation}
Assuming non-degeneracy, the only nontrivial way\footnote{The trivial case is 
$\hat P \ket{\kappa}=0$.} 
to achieve this is by matching $E_{\kappa}$ by one of the
eigenvalues of $\hat P \hat H \hat P$ and satisfying the zero-overlap condition for all
other eigenvectors:
\begin{equation}	 
\varepsilon_{\ell} = E_{\kappa} \; \land  \;  \braket{\varphi_{\ell'}}{\kappa} =0 \;\; \forall \;\, \ell'\neq \ell.
\label{eq:zeroo2}
\end{equation}
Let us inspect the right-hand side of Eq.~(\ref{eq:pse}) to understand how this can arise.
The obvious option is that $\hat P$ and $\hat H$ commute,
which is the case when $\cal P$-space is spanned by eigenstates of $\hat H$. The
more interesting situation is the following:
Let us use Eq.~(\ref{eq:pspace1}) and act with the Hamiltonian to the left:
\begin{equation}	 
\hat P \hat H \hat Q \ket{\kappa} = \sum_{j=1}^N \ket{\psi_j}\bra{\psi_j}\hat H \hat Q \ket{\kappa}
=  \sum_{j=1}^N \braket{\chi_j^Q}{\kappa} \ket{\psi_j} ,
\label{eq:phq}
\end{equation}
where we have defined
\begin{equation}	 
\ket{\chi_j^Q} = \hat Q \hat H \ket{\psi_j} .
\label{eq:chiq}
\end{equation}
For expression (\ref{eq:phq}) to be zero for a given $\ket{\kappa}$, 
we need $\braket{\chi_j^Q}{\kappa} = 0$ for $j=1, \ldots ,N$.
This does not seem achievable in general except if the image space of $\hat Q \hat H \hat P$ is
one-dimensional in which case there is only one such condition.
If $\ket{\kappa}$ lies in a continuum we can express this requirement as 
\begin{equation}
\chi^Q(\kappa) \equiv  \braket{\kappa}{\chi^Q} =0,
\label{eq:chik0a}
\end{equation}
where the single residual function $\chi^Q(\kappa)$ can be written as
\begin{eqnarray}
\chi^Q(\kappa) &=& \alpha_{\ell} \bra{\kappa}\hat Q \hat H \hat P \ket{\varphi_{\ell}} =
\alpha_{\ell} (\bra{\kappa} \hat H \ket{\varphi_{\ell}} - \bra{\kappa} \hat P \hat H \hat P \ket{\varphi_{\ell}}) \nonumber \\ 
&=& \alpha_{\ell} (E(\kappa)-\varepsilon_{\ell} ) \varphi_{\ell}(\kappa) , \hskip20pt \ell=1,\ldots , N ,
\label{eq:chik0b}
\end{eqnarray}
with state-specific constants $\alpha_{\ell}$ and $\varphi_{\ell}(\kappa) = \braket{\kappa}{\varphi_{\ell}}$. 
Equation~(\ref{eq:chik0a}) is trivially fulfilled for states 
$\ket{\varphi_{\ell'}}$ which happen to be eigenstates of both $\hat P \hat H \hat P$ and 
$\hat H$.\footnote{%
These states $\ket{\varphi_{\ell'}}$ are elements of the kernel of $\hat Q \hat H \hat P$.}
For all other $\hat P \hat H \hat P$ eigenstates we can infer from Eq.~(\ref{eq:chik0b})
that the zeros (\ref{eq:chik0a}) of the residual function 
determine the eigenvalues $\varepsilon_{\ell}$
and force the zero-overlap conditions as per Eq.~(\ref{eq:zeroo2}).

One way to obtain a one-dimensional image space of $\hat Q \hat H \hat P$ is to start from a
complete $L^2$ Hilbert space basis with respect to which the Hamiltonian matrix is of
tridiagonal form. 
This matrix structure is the key feature behind the $J$-matrix method~\cite{heller74, jmatrix}.  
If for arbitrary $N$ all basis states $\{\ket{\psi_j},j=1,\ldots,N\}$ are included in $\cal P$-space, 
it is readily shown that
\begin{equation}
\ket{\chi_j^Q} = 
	\bra{\psi_{N+1}}\hat H \ket{\psi_N} \delta_{j,N} \, \ket{\psi_{N+1}}  ,
\label{eq:chiqtri}
\end{equation}
i.e., the only nonzero residual state is directly proportional to the first
basis state that lies in $\cal Q$-space, and the zero-overlap condition is
satisfied as per the argument established above.

If the image  space of $\hat Q \hat H \hat P$ is multi-dimensional one may introduce the
(squared) norm of the state~(\ref{eq:phq}):
\begin{equation}
\Lambda (\kappa) := \bra{\kappa} \hat Q \hat H \hat P \hat H  \hat Q \ket{\kappa} = \sum_{j=1}^N |\chi_j^Q(\kappa)|^2 .
\label{eq:lambda1}
\end{equation}
Using similar arguments as in Eq.~(\ref{eq:chik0b}) we can cast the squared norm into the form
\begin{equation}
\Lambda (\kappa) = \sum_{\ell=1}^N (E(\kappa)-\varepsilon_{\ell} )^2 |\varphi_{\ell}(\kappa)|^2 . 
\label{eq:lambda2}
\end{equation}
A stricly positive $\Lambda(\kappa)$ is equivalent to the statement that the
zero-overlap condition~(\ref{eq:zeroo2}) does not hold, indicating 
that one can use $\Lambda(\kappa)$ as a test for a given basis 
by calculating it using Eq.~(\ref{eq:lambda1}) and checking it for zeros.

\section{The one-dimensional free-particle problem}
\label{sec:free}
Let us consider the free-particle problem in a one-dimensional (1D) world as a simple 
application. The spectrum of the Hamiltonian
\begin{equation}
	\hat H = \frac{\hat p^2}{2}
\label{eq:hfree}
\end{equation}
is continuous with eigenvalues $E(\kappa)=\kappa^2/2$ and admits eigenfunctions of plane-wave form
in position space:
\begin{equation}
	\phi_{\kappa}(x) = \braket{x}{\kappa} = \frac{1}{\sqrt{2 \pi}} \exp[i\kappa x], \hskip20pt
	\braket{\kappa}{\kappa '} = \delta (\kappa -\kappa ').
\end{equation}

\subsection{Harmonic oscillator basis}
\label{sec:osci}
As a first example of a $\cal P$-space basis we choose eigenstates of a harmonic oscillator
of frequency $\omega$.
When acted upon by the free-particle Hamiltonian (\ref{eq:hfree}) these states
transform according to a tridiagonal pattern~\cite{CTDL}:
\begin{equation}
\frac{\hat p^2}{2} \ket{\psi_j} = -\frac{\omega}{4} \Big(\sqrt{(j+1)(j+2)}\, 
\ket{\psi_{j+2}} - (2j+1) \ket{\psi_j} + \sqrt{j(j-1)}\,\ket{\psi_{j-2}}\Big) .
\label{eq:oscicoup}
\end{equation}
For a $\cal P$-space spanned by the first $N$ even-parity states
$\{\ket{\psi_j},j=0,2,\ldots , 2N-2  \}$ only the highest-lying
state couples to $\cal Q$-space, i.e., there is only
one non-zero state (\ref{eq:chiq})
\begin{equation}
\ket{\chi_j^Q}= - \frac{\omega}{4} \sqrt{2N(2N-1)} \,
	\ket{\psi_{2N}} \delta_{j,2N-2} \, .
\label{eq:chiqosci}
\end{equation}
This result is consistent with Eq.~(\ref{eq:chiqtri}), as it should be given that
the harmonic oscillator basis is complete and the Hamiltonian matrix is of
tridiagonal form [cf.~Eq.~(\ref{eq:oscicoup})].

The momentum-space representation of the residual state (\ref{eq:chiqosci}) is
\begin{equation}
 \chi_{2N-2}^Q(\kappa)= - \frac{\omega}{4} \sqrt{2N(2N-1)}\, \psi_{2N}(\kappa) 
\label{eq:chiqk}
\end{equation}
with
\begin{equation}
\psi_{2N}(\kappa) = \Big(\frac{\omega}{\pi}\Big)^{1/4} 
	\frac{(-i)^{2N}}{\sqrt{2^{2N} (2N)!\,\omega}} H_{2N}(\kappa/\sqrt{\omega}) 
	\exp[-\kappa^2/(2\omega)] ,
\label{eq:psi2nk}	
\end{equation}
where $H_{2N}$ are Hermite polynomials. The (even) oscillator 
eigenfunction~(\ref{eq:psi2nk}) has 
$N$ zeros at $\kappa>0$. According to Eq.~(\ref{eq:chik0b}) these
zeros $\{\kappa_1 , \ldots , \kappa_N\}$ are linked to the $N$ (positive) 
eigenvalues of $\hat P \hat H \hat P$
via $\epsilon_{\ell} = \kappa_{\ell}^2/2$, and the zero-overlap condition~(\ref{eq:zeroo2})
holds at each $\kappa_{\ell}$. 

\begin{figure}
\begin{center}
\resizebox{0.7\textwidth}{!}{\includegraphics{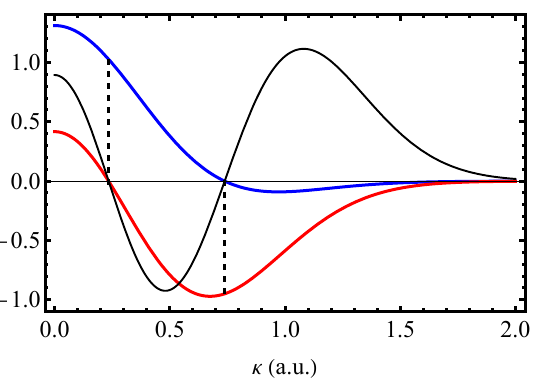}}
\vskip -0.5 truecm
\caption{%
Arbitrarily scaled residual function (\ref{eq:chiqk}) (thin black curve) for $N=2$ and eigenfunctions~(\ref{eq:osci-ef1}) (blue curve)
and~(\ref{eq:osci-ef2}) (red curve) for $\omega=0.2$ plotted in dependence of momentum $\kappa$. 
The vertical bars locate the momenta
$\kappa_{1,2}=\sqrt{2\varepsilon_{1,2}}$ that correspond to the matrix eigenvalues~(\ref{eq:secosc2}).
}
\label{fig:osci1}   
\end{center}
\end{figure}

Let us exemplify this for $N=2$.
The polynomial $H_4(\xi)=16\xi^4-48\xi^2+12$ has zeros at $\xi_{1,3}=\pm \sqrt{(3-\sqrt{6})/2}$
and $\xi_{2,4}=\pm \sqrt{(3+\sqrt{6})/2}$. 
Applying
Eq.~(\ref{eq:chik0b}) we infer that $\varepsilon_1 = \kappa_1^2/2 = \omega \,\xi_1^2/2 = \omega (3-\sqrt{6})/4 $, 
$\varphi_2(\kappa_1)=0$, $\varepsilon_2 = \kappa_2^2/2 = \omega \, \xi_2^2/2 = \omega (3+\sqrt{6})/4 $, and
$\varphi_1(\kappa_2)=0$. 
These results can be checked by diagonalizing the free-particle 
Hamiltonian~(\ref{eq:hfree})
in the two-dimensional $\cal P$-space spanned by $\ket{\psi_0}$ and
$\ket{\psi_2}$. The roots of the equation
\begin{equation}
\mbox{\rm det}(H-\varepsilon_{\ell} I) = 
\left| \begin{array}{cc}
H_{11}-\varepsilon_{\ell} & H_{12} \\
H_{21} & H_{22}-\varepsilon_{\ell}
\end{array} \right| = 0
\label{eq:secosci}	
\end{equation}
with $H_{11}=\omega/4$, $H_{12}=H_{21}=-\omega/(2\sqrt{2})$,
and $H_{22}=5\,\omega/4$ are
\begin{equation}
\varepsilon_{1,2}= \frac{\omega}{4} (3\mp\sqrt{6}) ,
\label{eq:secosc2}	
\end{equation}
as required, and the (normalized) eigenfunctions in momentum space read
\begin{eqnarray}
\varphi_{1}(\kappa) &=& \frac{1}{\sqrt{\varepsilon_2-\varepsilon_1}} \Big(\sqrt{H_{22}-\varepsilon_1} \, \psi_0(\kappa) 
+ \sqrt{H_{11}-\varepsilon_1} \, \psi_2(\kappa)\Big) , 
\label{eq:osci-ef1}	\\
\varphi_{2}(\kappa) &=& \frac{-1}{\sqrt{\varepsilon_2-\varepsilon_1}} \Big(\sqrt{H_{11}-\varepsilon_1} \, \psi_0(\kappa) 
- \sqrt{H_{22}-\varepsilon_1} \,\psi_2(\kappa) \Big) .
\label{eq:osci-ef2}
\end{eqnarray}
Figure~\ref{fig:osci1} shows the (arbitrarily scaled) residual function $\chi_2^Q(\kappa)$ [Eq.~(\ref{eq:chiqk})] 
together with the eigenfunctions~(\ref{eq:osci-ef1}) and (\ref{eq:osci-ef2}) for $\omega=0.2$.
As can be seen, the zero-overlap condition is satisfied at both matrix eigenvalues.

\begin{figure}
\begin{center}
\resizebox{0.7\textwidth}{!}{\includegraphics{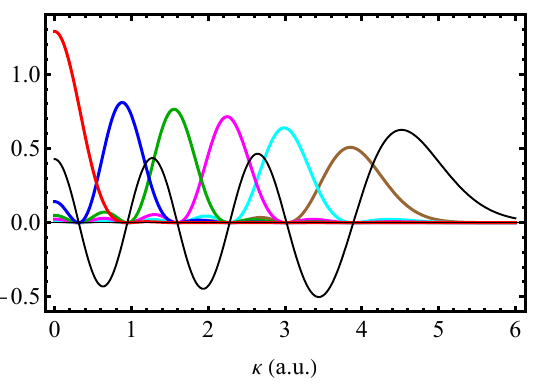}}
\vskip -0.5 truecm
\caption{%
Arbitrarily scaled residual function (\ref{eq:chiqk}) (thin black curve) for $N=6$ and squared eigenfunctions 
$\varphi_{\ell}^2(\kappa)$ ($\ell=1,\ldots, 6)$ (colored curves) for $\omega=0.2$ plotted in dependence 
of momentum $\kappa$.
}
\label{fig:osci2}   
\end{center}
\end{figure}

Figure~\ref{fig:osci2} provides a slightly different plot for $N=6$. To bring out more clearly  the `clumpy' structure
of the pseudostates referred to by McGovern \textit{et al.} in Ref.~\cite{mcgovern09}, we show their squares
together with the scaled residual function. Each distribution function $\varphi_{\ell}^2(\kappa)$ peaks close to the 
momentum $\kappa_{\ell}=\sqrt{2\varepsilon_{\ell}}$ and contributes mostly in the interval between the two adjacent
momenta $\kappa_{\ell-1}$  and $\kappa_{\ell+1}$. Small side maxima outside of this interval are, however, visible, 
indicating that
a one-to-one association of a $\hat P \hat H \hat P$ eigenstate with a momentum value (or energy) is only possible at
the matrix eigenvalues at which the zero-overlap condition occurs.
This has ramifications for the calculation of differential ionization probabilities in scattering calculations, as
was discussed for the Coulomb problem in Ref.~\cite{jj26}.

We can also use this simple example to illustrate the case of a multi-dimensional
image space of $\hat Q \hat H \hat P$ for which the zero-overlap condition is violated.
Removing the oscillator ground state from $\cal P$-space creates a second residual function
[cf.~Eq.~(\ref{eq:oscicoup})]
\begin{equation}
	\chi_{2}^Q(\kappa)= - \frac{\omega}{4} \sqrt{2} \, \psi_{0}(\kappa) 
\label{eq:chiq2}
\end{equation}
so that the squared norm~(\ref{eq:lambda1}) becomes
\begin{equation}
	\Lambda (\kappa) = \frac{\omega^2}{16} \Big(2 \psi_0^2 (\kappa) + 2N(2N-1) \psi_{2N}^2(\kappa) \Big) .
\label{eq:lambda_osci}
\end{equation}
This function does not have zeros since the ground-state oscillator wave function 
$\psi_0(\kappa)$ is nodeless. The zero-overlap condition does not hold in this case.

\subsection{Momentum-space basis functions}
\label{sec:pspace}
As a second example 
we consider a $\cal P$-space spanned by two simple even-parity momentum-space functions
\begin{equation}
	v_1(\kappa) = \frac{1}{(\kappa^2 +1)^2} , \hskip10pt
        v_2(\kappa) = \frac{\kappa^2}{(\kappa^2 +1)^2} ,
\label{eq:pspace}
\end{equation}
which are neither normalized nor orthogonal to each other. 
Operating with the 1D free-particle Hamiltonian (\ref{eq:hfree}) on the first
function produces the second function (multiplied by 1/2), while
\begin{equation}
	\frac{\hat p^2}{2} v_2(\kappa) = \frac{\kappa^4}{2(\kappa^2 +1)^2}
	\label{eq:p2v2}
\end{equation}
cannot be written as a linear combination of $v_1(\kappa)$ and $v_2(\kappa)$.
This is to say that only one linearly-independent residual function $\chi^Q(\kappa)$ exists, i.e., the
image space of $\hat Q \hat H \hat P=(\hat 1 - \hat P)\hat H \hat P$ is one-dimensional.
Orthogonalizing (\ref{eq:p2v2}) to $\cal P$-space yields (with arbitrary scaling applied)
\begin{equation}
	\chi^Q(\kappa) =  \frac{(1-6\kappa^2+\kappa^4)}{(\kappa^2 +1)^2} .
\label{eq:chiqp2}	
\end{equation}
As per Eq.~(\ref{eq:chik0b}), the zeros 
of the even function~(\ref{eq:chiqp2}) 
determine the matrix eigenvalues $\varepsilon_{1,2}=3/2\mp \sqrt{2}$ of the
Hamiltonian in the two-dimensional basis (\ref{eq:pspace}) and force the zero-overlap condition~(\ref{eq:zeroo2}) 
to hold at the corresponding set of momenta $\kappa_{\ell}$. 

We note that at no point in this analysis was it necessary to determine a single $\cal Q$-space function, let alone the complete set of functions 
that span Hilbert space when combined with the two $\cal P$-space functions~(\ref{eq:pspace}).
On the other hand, it appears that there is no straightforward way to increase 
the set~(\ref{eq:pspace}) without violating the zero-overlap condition:
The function on the right-hand side of Eq.~(\ref{eq:p2v2}) may look like the obvious choice for a third
basis state but is not square-integrable. Turning it into an $L^2$ function by, e.g., 
increasing the exponent in the denominator and adding
\begin{equation}
	v_3(\kappa) = \frac{\kappa^4}{(\kappa^2 +1)^4}
	\label{eq:p2v3}
\end{equation}
to $\cal P$-space results in a second, linearly-independent, residual function, i.e., a
two-dimensional $\hat Q \hat H \hat P$  image space, and a
squared norm~(\ref{eq:lambda1}) which does not have zeros.

As a final point we note that the Hamiltonian matrix with respect to a Gram-Schmidt-orthogonalized 
version of the basis set $\{v_1(\kappa),v_2(\kappa),v_3(\kappa)\}$ turns out to be of tridiagonal
form. This is not a contradiction to the statement made in Sec.~\ref{sec:theory} since Eq.~(\ref{eq:chiqtri}) 
does not apply to the present case
in which we have no explicit knowledge of the $\cal Q$-space basis.
We can conclude that tridiagonality of $\hat P \hat H \hat P$ in itself is not a sufficient criterion for
the zero-overlap condition to hold.

\section{The Coulomb problem} 
\label{sec:coulomb}

The Coulomb Hamiltonian in coordinate-space representation reads
\begin{equation}
	\hat H = -\frac{1}{2} \nabla^2 - \frac{1}{r} .
\label{eq:hcoul}
\end{equation}
For non-negative energies its spectrum is continuous and we can write
$E(\kappa)=\kappa^2/2$ where $\kappa$ is the wave number, i.e., the magnitude of the 
wave vector $\pmb{\kappa}$.
The eigenfunctions are usually referred to as Coulomb waves and
can be summarized as~\cite{joachain}
\begin{equation}
	\phi_{\pmb{\kappa}} (\mathbf{r}) = \sqrt{\frac{2}{\pi}} \sum_{lm} i^l e^{i\sigma_l}
	\frac{F_l(\kappa r)}{\kappa r} Y_{lm}^*(\Omega_\kappa) Y_{lm}(\Omega_r),
\label{eq:cws}
\end{equation}
where the $F_l$ are the solutions of the radial Coulomb wave equation, 
$\sigma_l = {\rm arg}[\Gamma(l+1+i\eta)]$
is the Coulomb phase shift, $\eta = -1/\kappa$ is the Sommerfeld parameter, and $Y_{lm}$
are spherical harmonics.

We are interested in discretizing the Coulomb continuum in terms of a 
finite set of square-integrable functions. That is, we
consider basis states of the form
\begin{equation}
	\psi_{klm} (\mathbf{r}) = \frac{\zeta_{kl}(r)}{r} Y_{lm}(\Omega_r)
\label{eq:laguerre1}
\end{equation}
with 
(real) radial $L^2$ functions $\zeta_{kl}(r)$.
For the ensuing analysis, it suffices 
to consider the radial components of the residual functions that correspond
to the states~(\ref{eq:chiq}). In a first step, we construct them in position
space:
\begin{equation}
	\chi_{kl}^Q(r) = (1- {\hat P}^{(l)}) \hat H_r^{(l)} \zeta_{kl}(r) ,
\label{eq:chiqr}
\end{equation}
where 
\begin{equation}
\hat H_r^{(l)} = - \frac{1}{2} \frac{d^2}{dr^2} + \frac{l(l+1)}{2r^2} - \frac{1}{r}
\label{eq:hrad}
\end{equation}	
is the radial Hamiltonian for a given angular momentum quantum number $l$ and 
\begin{equation}
{\hat P}^{(l)} f(r) = \sum_{k=1}^{N_l} \zeta_{kl}(r) \int_0^{\infty} \zeta_{kl}(r') f(r') dr'
\label{eq:plfr}
\end{equation}
is the projection of the function $f(r)$ onto the set $\{\zeta_{kl}(r),k=1,\ldots N_l\}$.
Note that in order for Eq.~(\ref{eq:plfr}) to be correct the basis functions
need to be normalized and mutually orthogonal.
In a second step, we generate the residual functions in $\kappa$-space by explicit projection
onto the radial Coulomb functions in Eq.~(\ref{eq:cws}).
We have done this numerically in the analysis that follows. 

We consider a set of orthonormal Laguerre functions which for each angular momentum quantum number $l$ can be
written as~\cite{bray17}
\begin{equation}
\zeta_{kl}(r)
	= {\cal N}_{kl} x_l^{l+1} L_{k-1}^{2l+2}(x_l) e^{-x_l/2} , \hskip10pt 
k=1, \ldots , N_l
\label{eq:laguerre2}
\end{equation}
with the normalization constants
${\cal N}_{kl}=\sqrt{(\lambda_l (k-1)!)/(2l+k+1)!}$, the
associated Laguerre polynomials $L_{k-1}^{2l+2}$, free parameters $\lambda_l$, and
scaled coordinates $x_l=\lambda_l r$. 
We show in the Appendix that for this choice of basis one can work out the right-hand side
of Eq.~(\ref{eq:chiqr}) to find
\begin{equation}
\chi_{kl}^{Q}(r)= \alpha_{kl} x_l^{l} L_{N_l}^{2l+1}(x_l) e^{-x_l/2} ,
\label{eq:laguerre-res}
\end{equation}
i.e., one linearly-independent residual function for each $l$ with state-specific 
constants $\alpha_{kl}$.\footnote{Note the different upper index in the associated
Laguerre polynomial in Eq.~(\ref{eq:laguerre-res}) compared to the basis functions~(\ref{eq:laguerre2}).} 
In other words, for each $l$ the image space of $\hat Q \hat H \hat P$ is one-dimensional,
and condition~(\ref{eq:chik0a}) can be analyzed to determine the (positive) energy
eigenvalues of $\hat P \hat H \hat P$ and to confirm the zero-overlap conditions using
Eq.~(\ref{eq:chik0b}). 
We note in passing that the same function~(\ref{eq:laguerre-res}), albeit with slightly different
coefficients $\alpha_{kl}$, is obtained as residual function if the Coulomb potential is omitted
from the Hamiltonian~(\ref{eq:hrad}), i.e., if the three-dimensional free-particle problem is
considered instead of the Coulomb problem. This can be inferred without difficulty from the
details provided in the Appendix.

\begin{figure}
\begin{center}
\resizebox{0.7\textwidth}{!}{\includegraphics{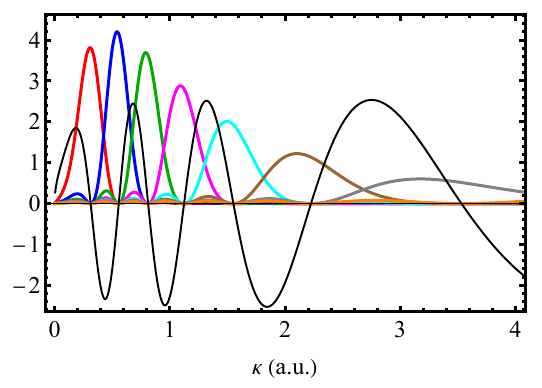}}
\vskip -0.5 truecm
\caption{%
Arbitrarily scaled residual function (\ref{eq:laguerre-res}) (thin black curve) and squared eigenfunctions
obtained from diagonalizing the radial Hamiltonian~(\ref{eq:hrad})
in a Laguerre basis for $l=0$, $\lambda_0=2$, and $N_0=12$ (colored curves) 
plotted in dependence of the wave number $\kappa$.
All (eight) eigenfunctions for positive energy eigenvalues are included, but only seven of them are
clearly visible since the largest
eigenvalue occurs at a $\kappa$-value outside of the interval displayed (see text for details).
}
\label{fig:lag}   
\end{center}
\end{figure}

As an example for the Coulomb problem, Fig.~\ref{fig:lag} shows in $\kappa$-space the 
residual function~(\ref{eq:laguerre-res}) 
for $l=0$, $\lambda_0=2$, and $N_0=12$. 
The function has eight zeros, one of which occurs outside of the wave number range shown at $\kappa\approx 7.43$. 
Accordingly, the diagonalization of the radial Hamiltonian~(\ref{eq:hrad}) in this twelve-dimensional basis 
yields four negative and eight positive eigenvalues, the latter matching exactly with the zeros of the
residual function.
The four negative \textit{energy} eigenvalues are
$-0.5, -0.125, -0.05497, -0.01517$, i.e., the ground state is included exactly for the choice
$\lambda_0=2$, the $2s$ eigenvalue is exact up to more than five significant figures, and the next two excited
$s$-states are represented with decreasing accuracy.

All (squared) eigenfunctions corresponding to the positive eigenvalues are also included in Fig.~\ref{fig:lag}, 
but only seven of them are clearly visible in the $\kappa$-range shown.
A similar plot of squared eigenfunctions in $\kappa$-space was provided by McGovern {\it et al.} 
in Fig.~1 of Ref.~\cite{mcgovern09}, albeit for a larger basis consisting of 21
Laguerre functions and a slightly smaller value for $\lambda_0$,
tuned such that the 10$^{\rm th}$ eigenvalue occurs exactly at an energy of 5~eV. 

The structure of these functions 
is similar to those of the 1D free-particle problem represented in the harmonic oscillator basis 
(cf.~Fig.~\ref{fig:osci2}), exhibiting pronounced
maxima close to the wave numbers that correspond to their eigenvalues, exact zeros at all other eigenvalues,
and smaller side maxima in between. 
Looking at both figures, and also at Fig.~1 of Ref.~\cite{mcgovern09}, the ``clumpiness'' appears to 
become more pronounced with
increasing $\cal P$-space dimension. The occurrence of exact zeros at the matrix eigenvalues, i.e., the zero-overlap 
condition~(\ref{eq:zeroo2}) is, however, independent of basis size, as demonstrated by the present analysis.

\section{Conclusions}
\label{sec:conclusions}
In this paper, we have presented an analysis of the zero-overlap condition, i.e., the previously observed
phenomenon that a set of pseudostates obtained from diagonalizing a Hamiltonian $\hat H$ with a (partially) continuous spectrum 
in an $L^2$ basis set may have exact zeros in the eigenrepresentation of $\hat H$
at all pseudo-continuum matrix eigenvalues 
except the one a given pseudostate is associated with as eigenvector.
Our analysis shows that this occurs if the image space of the operator $\hat Q \hat H \hat P$, where $\hat P$
is the projection operator onto the subspace spanned by the $L^2$ basis and $\hat Q = \hat 1 - \hat P$ its
complement, is one-dimensional. 
If this is the case, complete information on the pseudo-continuum matrix eigenvalues is encoded in one
function, called the residual function in this work, and can be extracted by determining its zeros in 
the eigenrepresentation of $\hat H$. While this might not offer a practical advantage compared to 
computing the eigenvalues via matrix diagonalization it is an interesting (and unexpected) result.

For the free-particle problem (in 1D), a basis of harmonic oscillator
eigenstates and a minimal set of momentum-space wave functions, and for the Coulomb problem (in 3D), a Laguerre basis set are shown to satisfy the (sufficient)
condition of a one-dimensional $\hat Q \hat H \hat P$ image space.
This provides a more general context of the ovservation made for the Laguerre basis in Ref.~\cite{mcgovern09} 
and proved in Ref.~\cite{ccc1} 
using rather specific properties of the basis functions. The present proof for the Laguerre basis appears
technical as well, but can be understood in more qualitative terms by recognizing that the application of the
Coulomb Hamiltonian on a basis function of the form $(c_1 r^{l+1} + c_2 r^{l+2} + \dots + c_k r^{l+k}) 
\exp[-\lambda_l r/2]$ generates the function 
 $(d_0 r^l + d_1 r^{l+1} + d_2 r^{l+2} + \dots + d_k r^{l+k}) \exp[-\lambda_l r/2]$. 
The first term of this function is the only one that is structurally different from the components of
the basis function,
indicating that the image space of $\hat Q \hat H \hat P$ is one-dimensional.

We demonstrated in Ref.~\cite{jj26} that the occurrence of the zero-overlap condition eliminates
spurious asymptotic channel couplings in time-dependent calculations in which a pseudostate-expanded 
state vector is projected onto true eigenstates of the asymptotic Hamiltonian to extract the transition
probabilities of interest, i.e., it guarantees the asymptotic stability of these results. This makes it
a desirable property of a basis set used in a coupled-channel calculation of ionization transitions in
particular.

A multi-dimensional image space of $\hat Q \hat H \hat P$ does not preclude the zero-overlap condition from being
met exactly but makes it extremely unlikely [cf.~Eqs.~(\ref{eq:phq}) and~(\ref{eq:lambda1})]. 
We have found numerically that basis sets that are associated with such image spaces may satisfy the
condition in good approximation. The BGM basis used in Ref.~\cite{jj26} for the calculation of
differential electron emission cross sections in antiproton--hydrogen collisions is one such example. 
Another basis which does not satisfy the zero-overlap condition exactly, but can get close to a one-dimensional
residual function space is the spherical Gauss-Laguerre basis.
We are currently
analyzing these and other widely used basis sets 
with respect to this criterion and will report on these studies elsewhere.

\begin{acknowledgments}
Financial support from the Natural Sciences and Engineering Research Council of Canada (Grants No. RGPIN-2023-05072 and No. RGPIN-2025-06277) 
is gratefully acknowledged. One of us (T. K.) thanks Jay Jay Tsui and Janakan Sivasubramanium for discussions.
\end{acknowledgments}

\appendix*
\section{}
Applying the radial Hamiltonian~(\ref{eq:hrad})
to the Laguerre functions (\ref{eq:laguerre2}) and using the recurrence relations
\begin{equation}
x L_k^{\alpha}\, '(x) = k L_k^{\alpha}(x) - (k+\alpha) L_{k-1}^{\alpha}(x)
\label{eq:lag1}
\end{equation}	
and
\begin{equation}
x L_{k-2}^{\alpha+1}(x) = -(k-1) L_{k-1}^{\alpha}(x) + (k+\alpha -1) L_{k-2}^{\alpha}(x) ,
\label{eq:lag2}
\end{equation}
together with the differential equation~\cite{GR}
\begin{equation}
x L_k^{\alpha}\, ''(x) + (\alpha +1-x) L_{k}^{\alpha}\, '(x) +k L_k ^{\alpha}(x) = 0
\label{eq:lag3}
\end{equation}	
yields 
\begin{eqnarray}
\hat H_r^{(l)} \zeta_{kl}(r) &=&  \frac{\lambda_l^2}{2} {\cal N}_{kl} e^{-x_l/2} 
\Big\{\!\! -x_l^{l} L_{k-2}^{2l+3} (x_l) + \Big(k+l-\frac{2}{\lambda_l}\Big) x_l^{l} L_{k-1}^{2l+2} (x_l) 
- \frac{1}{4} x_l^{l+1} L_{k-1}^{2l+2} (x_l) \Big\}  \nonumber \\
&=& \frac{\lambda_l^2}{2} {\cal N}_{kl} e^{-x_l/2} \Big\{\!\!
-x_l^{l} L_{k-2}^{2l+3} (x_l) + \Big(k+l-\frac{2}{\lambda_l}\Big) x_l^{l} L_{k-1}^{2l+2} (x_l)\Big\} 
	- \frac{\lambda_l^2}{8} \zeta_{kl}(r) 
\label{eq:lag4}
\end{eqnarray}
with the understanding that $L_{k-2}^{2l+3} (x_l)=0$ for $k=1$.
The last term on the right-hand side of~(\ref{eq:lag4}) obviously lies in ${\cal P}^{(l)}$-space,
the space spanned by the Laguerre basis functions~(\ref{eq:laguerre2}) for a given $l$. 
The other terms 
mostly lie in ${\cal P}^{(l)}$-space as well since the terms in curly brackets are polynomials
which for the most part can be represented by the polynomial terms in~(\ref{eq:laguerre2}).
The exception are the lowest-order terms arising from multiplying $x_l^l$ by the constant
terms in $L_{k-2}^{2l+3}(x_l)= c_{k-2}^{2l+3} x_l^0 + \ldots $ and 
$L_{k-1}^{2l+2}(x_l) = c_{k-1}^{2l+2} x_l^0 + \ldots$. This is to say that
\begin{eqnarray}
(1 - {\hat P}^{(l)} ) \Big(e^{-x_l/2} x_l^l  L_{k-2}^{2l+3}(x_l) \Big) 
&=& c_{k-2}^{2l+3} \Big(e^{-x_l/2} x_l^l - {\hat P}^{(l)} (e^{-x_l/2} x_l^l  ) \Big) , 
\label{eq:lag5} \\
  (1 - {\hat P}^{(l)} ) \Big(e^{-x_l/2} x_l^l  L_{k-1}^{2l+2}(x_l) \Big) 
&=& c_{k-1}^{2l+2} \Big(e^{-x_l/2} x_l^l - {\hat P}^{(l)} (e^{-x_l/2} x_l^l) \Big) .
\label{eq:lag6} 
\end{eqnarray}
The projections in (\ref{eq:lag5}) and (\ref{eq:lag6})
\begin{equation}
	{\hat P}^{(l)} (e^{-x_l/2} x_l^l) = \sum_{j=1}^{N_l} I_j^{(l)} \zeta_{jl}(r)
\label{eq:project1}
\end{equation}
with
\begin{equation}
	I_j^{(l)} = \int_0^{\infty} \zeta_{jl}(r) e^{-x_l/2} x_l^l \,  dr
\label{eq:project2}
\end{equation}
can be carried out using the integral~\cite{GR}
\begin{equation}
\int_0^{\infty} x_l^{2l+1} L_{j-1}^{2l+2}(x_l) e^{-x_l} dx_l  = (2l+1)!
\label{eq:projectint}
\end{equation}
to find the intermediate result
\begin{equation}
{\hat P}^{(l)} (e^{-x_l/2} x_l^l) = e^{-x_l/2} x_l^{l+1} (2l+1)!  
\sum_{j=1}^{N_l} \frac{(j-1)!}{(2l+j+1)!} L_{j-1}^{2l+2}(x_l) .
\label{eq:project3}
\end{equation}
Using once again the recurrence relation~(\ref{eq:lag2}) we can 
eliminate the sum over $j$ on the right-hand side of~(\ref{eq:project3})
by going through the following steps:
\begin{eqnarray}
(2l+1)! && \!\!\!\!\! \sum_{j=1}^{N_l} \frac{(j-1)!}{(2l+j+1)!} x_l L_{j-1}^{2l+2}(x_l) 
\nonumber \\
&=& (2l+1)! \Big(-\sum_{j=1}^{N_l} \frac{j!}{(2l+j+1)!} L_j^{2l+1}(x_l) 
+  \sum_{j=1}^{N_l} \frac{(j-1)!}{(2l+j)!} L_{j-1}^{2l+1}(x_l) \Big) 
\nonumber \\
&=& (2l+1)! \Big(-\sum_{j=1}^{N_l} \frac{j!}{(2l+j+1)!} L_j^{2l+1}(x_l) 
+ \frac{1}{(2l+1)!} +  \sum_{j=1}^{N_l-1} \frac{j!}{(2l+j+1)!} L_{j}^{2l+1}(x_l) \Big) 
\nonumber \\
&=& 1 - \frac{(2l+1)!N_l!}{(2l+N_l+1)!} L_{N_l}^{2l+1}(x_l) .
\label{eq:resum} 
\end{eqnarray}
We thus arrive at the compact form
\begin{equation}
{\hat P}^{(l)} (e^{-x_l/2} x_l^l) = e^{-x_l/2} x_l^{(l)} 
\Big( 1- \frac{(2l+1)!N_l!}{(2l+N_l+1)!} L_{N_l}^{2l+1}(x_l) \Big) ,
\label{eq:project4}
\end{equation}
and together with Eqs.~(\ref{eq:lag4})--(\ref{eq:lag6}) obtain for the residual function~(\ref{eq:chiqr})
the expression in Eq.~(\ref{eq:laguerre-res}) with
\begin{equation}
\alpha_{kl} = \frac{\lambda_l}{2} {\cal N}_{kl} \Big\{-c_{k-2}^{2l+3}
+\Big(k+l-\frac{2}{\lambda_l}\Big) c_{k-1}^{2l+2}\Big\} \frac{(2l+1)!N_l!}{(2l+N_l+1)!} .
\label{eq:alphal}
\end{equation}

If the free-particle problem is considered instead of the Coulomb problem the analysis remains
unchanged, except that the terms with the $2/\lambda_l$ prefactor are absent from Eqs.~(\ref{eq:lag4}) and~(\ref{eq:alphal}).

\bibliography{overlap}

\end{document}